# Enhanced critical current density of MgB$_2$ superconductor synthesized in high magnetic fields

Yanwei Ma [a], Aixia Xu, Xiaohang Li, Xianping Zhang

Applied Superconductivity Lab., Institute of Electrical Engineering,
Chinese Academy of Sciences, P. O. Box 2703, Beijing 100080, China

S. Awaji, K. Watanabe

High Field Laboratory for Superconducting Materials, Institute for Materials Research,
Tohoku University, Sendai 980-8577, Japan

**Abstract:**

The effect of high magnetic fields on the current carrying properties of both MgB$_2$ bulks and Fe-sheathed tapes was investigated following different thermal sequences. It is found that application of a large magnetic field during processing results in the quite uniform microstructure and the better connectivity between the MgB$_2$ grains. As a result, the Jc of these samples has shown much higher value than that of the MgB$_2$ samples in the absence of magnetic field. The possible mechanism of the Jc enhancement under an external magnetic field is also discussed.

[a] Electronic mail: ywma@mail.iee.ac.cn

Magnesium diboride ($MgB_2$) with its superconducting transition temperature at 39 K has generated much interest in promising applications potential in high field magnets and magnetic resonance imaging [1]. At present, research efforts have been directed towards either improving critical current density (Jc) by improving grain connectivity, or improving in-field performance through introducing pinning centers. To achieve such goals, doping with elements or compounds [2-7], hot isostatic pressing [8], and irradiation with heavy ions [9] has been investigated.

Processing in an external magnetic field is a well-proven technique to enhance the degree of grain alignment and critical current density for the case of high-Tc oxide superconductors (HTS) [10-12]. The mechanism operating to achieve such improvement is understood to be due to the anisotropy in the paramagnetic susceptibility. Hence, when high-Tc superconductors are placed in a magnetic field in its normal state, the magnetic energy is minimized when the axis of maximum susceptibility is parallel to the magnetic field. On the other hand, $MgB_2$ exhibits a strong anisotropy in the B-B lengths: the distance between the boron planes (c-axis) is significantly longer than the in plane B-B distance (a axis) [1]. Furthermore, it is known that the Jc properties of $MgB_2$ are quite sensitive to preparation and annealing conditions. If a magnetic field is applied during the $MgB_2$ fabrication process, the enhancement of critical current density and other field effects are expected. In this letter, we have found that the magnetic field sintering (MFS) effectively improves the current carrying properties of both $MgB_2$ bulks and Fe-clad tapes.

Powders of Mg (99.8%, 325 mesh) and B (amorphous, 99.99%) were well mixed and ground in air for 1 h, using an agate mortar and pestle. Pellets 10 mm in diameter and 2 mm in thickness were made under uniaxial pressure. Fe-sheathed $MgB_2$ tapes with a thickness of ~ 0.5 mm and a width of ~ 3.5 mm were prepared by the standard powder-in-tube method. Subsequently, the pellet or tape samples were wrapped in Zr foil, placed in a vertical tube furnace, then heated in vacuum in applied magnetic fields ($H_a$) up to 14 T following different thermal sequences. As for the tapes, several groups of samples were prepared in this experiment, based on that the surface of the tapes was oriented whether parallel to or perpendicular to the magnetic field. A description of the samples is presented in table 1.

The samples were sintered in an electrical furnace in a vacuum of about $10^{-4}$ Pa. The furnace was installed in a 15 T cryogen-free type superconducting magnet. For the samples sintered in a magnetic field, first, the magnetic field was raised up to the set value; second, the temperature was heated up to the target value and the samples were sintered at the set temperature in a magnetic field for 1h; then the samples were furnace-cooled down to room temperature; finally the magnetic field was decreased to zero. $MgB_2$ samples were also prepared in the absence of a magnetic field and used as the standard. Several different samples were made in separate reaction runs to check for reproducibility.

Phase identification was performed by x-ray diffraction (XRD) using Cu Kα radiation. For study of tapes, the Fe sheaths were mechanically removed by peeling off the sheath to expose the core. Microstructural observation was carried out by scanning electron microscopy (SEM). Transport critical current densities (Jc) of some tape samples were measured at 4.2 K using a conventional four-probe method. The criterion for the Ic definition was 1 µV/cm. A magnetic field was applied parallel to the tape surface. The hysteretic magnetization ΔM of samples was also measured in a superconducting quantum interference device magnetometer, from which the critical current density Jc was calculated assuming fully connected samples using the extended bean model: Jc=20ΔM/[$a$(1-$a$/3$b$)] where $a$ and $b$ are the width and thickness of a rectangular section bar.

Figure 1 shows the x-ray diffraction patterns of the superconducting cores of Fe sheathed tapes processed in magnetic fields $H_a$ of 0 and 10 T at 600°C for 1h (Group I). As we can see, both the zero field and 10 T samples compose of almost a single phase of $MgB_2$ containing a small amount of MgO. However, the relative intensities of the (001) and (002) diffraction peaks from $MgB_2$ for the 10T tapes are lower than those of the 0T samples. These results indicate that the c-axis grain alignment of $MgB_2$ was deteriorated by the magnetic field sintering.

Figure 2 shows the transport critical current densities at 4.2 K as a function of magnetic fields for our 0 and 10 T processing $MgB_2$/Fe tapes. The tapes processed in a 10 T magnetic field exhibit higher Jc values than those processed under a zero magnetic field, although the Jc difference is small in magnetic fields below 10 T. In particular, it is evident that the field dependence of Jc was decreased by sintering in a magnetic field for Fe sheathed tapes, namely, higher Jc in high-field region. Note that the similar result was found for the tapes made in a 14 T field, when sample's surface was placed parallel to the field direction (not shown in Table 1). Furthermore, magnetization data reveal that compared to the 0 T tape, the critical temperature Tc for the samples processed in a 10 T magnetic field slightly decreased by 0.3 K. The small decrease of Tc is likely due to the slightly poor crystallinity originating from magnetic field processing, as supported by weaken XRD patterns with broad peaks.

To clarify the influence of field direction on the tape surface during MFS processing, we set horizontally the Fe sheath tapes in the furnace and obtained the Group II samples. Figure 3 shows the magnetic Jc(B) curves for the $MgB_2$ tapes processed in magnetic fields $H_a$ of 0 and 14 T at 700°C for 1h. It is noted that the tapes processed in a 14 T magnetic field has better Jc(B) performance, more than two times higher than the 0T ones. In order to investigate the reason for the Jc improvement, we studied the difference in the microstructure of the tapes with and without magnetic field. Figure 4 shows the typical SEM images of the fractured core layers for the 0T and 14 T samples. Clearly, well-developed grains can be seen in both samples. However, the 0T sample shows a broader grain size distribution and the core is quite porous and loose (see Fig.4 (a)). The porous microstructure of MgB2 directly means a reduction of the effective current path. In contrast, with the application of strong magnetic field, the field sample has fewer pores and seems very dense and consequently the connections between grains are much improved. Moreover, the quite uniform microstructure of the $MgB_2$ core is observed (see Fig.4 (b)). As were demonstrated earlier [14], the high-density $MgB_2$ samples with less voids have high superconducting homogeneity and strong intergranular current flow as determined by magneto-optical studies. The fact was also corroborated by many recent results, in which the Jc enhancement of $MgB_2$ was achieved by the improvement in the grain coupling as a consequence of densification of the tape core [3,7-8].

It is interesting to note by comparing Figs.1 and 3, the effect of a magnetic field seems different between Group I and II samples. For Group I (the applied field was in the tape plane), although the enhanced Jc-B characteristic was observed in high field region, however, the Jc improvement in low field area is small. On the other hand, the improved Jc by more than a factor of 2 for the field tapes of Group II was achieved. This indicates that the magnetic field works more effectively to enhance the Jc-B properties when the direction of applied fields was perpendicular to the tape surface (Group II) during processing.

To further confirm the enhanced Jc properties with the magnetic field, we also prepared the $MgB_2$ bulk samples using the *in situ* reaction in a magnetic field 800°C for 1h. Figure 5 presents the magnetic Jc(B) data at 5, 20 and 30 K for the $MgB_2$ pellets processed in magnetic fields $H_a$ of 0 and 8 T. It is noted that the 8 T field produced a stronger enhancement of Jc than without the field at all temperatures and in the entire field region. The 8 T bulks show the slightly degraded Tc compared to the 0T ones, as given in the table 1. From the SEM observations for all bulk samples (Figs.4 (c) and (d)), it is clear that the microstructure development with the magnetic field is consistent with the tape case. Thus, we conclude that when the $MgB_2$ pellets processed in a magnetic field, Jc is also significantly improved due to the denser microstructure and the improved connectivity of $MgB_2$ grains. More details about bulks heat-treated in a magnetic field will be published elsewhere. Combined with the results of bulk and tape samples sintering in magnetic fields, our present study demonstrates that the magnetic field does enhance the Jc properties in $MgB_2$ significantly.

To understand the mechanism behind the Jc improvement, one may consider first the crystal orientation effect of $MgB_2$ during the magnetic field processing, since like HTS, the structure of $MgB_2$ is also strongly anisotropic. In the process of crystal growth in a magnetic field, if the anisotropic magnetic energy of crystal $\Delta E=\Delta\chi V H_a^2/2$, which promotes crystal alignment magnetically, exceeds the thermal energy, the grain alignment is obtained [10], where V is the volume of a grain and $\Delta\chi$ is the anisotropy of the paramagnetic susceptibility. Thus, it is expected that c-axis grain alignment of the $MgB_2$ core might be preferred by the strong magnetic field imposed. However, studies of the magnetism of the normal state of $MgB_2$ show that the net susceptibility is very small, of the order of $10^{-6}$ emu/mol [15]. Furthermore, the average grain size of $MgB_2$ in the present work is quite small (about 200 nm), while the HTS has the grains with a size of ~20-50 μm [12]. These suggest that the $MgB_2$ crystal orientation effect due to the applied magnetic field would be tiny, compared to HTS superconductors. Other beneficial effects such as densification and homogenization of crystallites caused by Lorentz force and magnetization force must be considered [16-17]. Therefore, the clear Jc enhancement in $MgB_2$ is mainly due to the well-connected grains of uniform size as a consequence of densification of the core during magnetic sintering.

As revealed by microstructural analyses, it is evident that an external magnetic field can enhance the densification and the homogeneity of superconducting cores. In particular, the magnetic field effect seems to be of great significance for group II when the applied field was perpendicular to the tape plane. We can see that the density of pores in the magnetically sintered sample are much less than that in the 0T one. Similar results were also found by Tsurekawa et al. [18] when sintering iron samples in the presence of a magnetic field. This is probably because a magnetic field provides an extra driving force for grain boundary migration to break away the dragging pores. Another interesting feature is that the grain structure seems to be homogenous, suggesting the abnormal grain growth was effectively prevented by a magnetic field [19]. Therefore, the applied magnetic field may provide a driving force for grain boundary migration that greatly contributes to densification during sintering, leading to better connections between grains, hence the Jc enhancement.

In summary, $MgB_2$ bulks and tapes were prepared in high magnetic fields up to 14 T. Enhanced Jc properties in comparison with their zero-field counterpart were observed. Application of a large magnetic field during processing produced the uniform and dense microstructure in $MgB_2$, leading to increased critical current density Jc. It is suggested that an external magnetic field is responsible for an increase in the driving force for grain boundary migration that greatly contributes to densification. The present study demonstrated that the magnetic field is very effective on yielding $MgB_2$ superconductors with the enhanced Jc properties.

The authors would like to thank Prof. K. Togano, P. Badica, T. Nishizaki, T. Nojima in IMR, Tohoku Univ. and Yulei Jiao, Ling Xiao in GRINM, Beijing for their help during the experiments. This work is partially supported by the National Science Foundation of China under Grant No.50472063 and No.50572104.

# References

[1] J. Nagamatsu, N. Nakagawa, T. Muranaka, Y. Zenitani, and J. Akimitsu, Nature (London) 410, 63 (2001).

[2] Y. Feng, Y. Zhao, Y. P. Sun, F. C. Liu, Q. Fu, L. Zhou, C. H. Cheng, N. Koshizuka, and M. Murakami, Appl. Phys. Lett. 79, 3983 (2001).

[3] Yanwei Ma, H. Kumakura, A. Matsumoto, and K. Togano, Appl. Phys. Lett. 83, 1181 (2003).

[4] J. Wang, Y. Bugoslavsky, A. Berenov, L. Cowey, A. D. Caplin, L. F. Cohen, J. L. MacManus-Driscoll, L. D. Cooley, X. Song, and D. C. Larbalestier, Appl. Phys. Lett. 81, 2026 (2002).

[5] H. Kumakura , H. Kitaguchi, A. Matsumoto and H. Hatakeyama, Appl. Phys. Lett. 84, 3669 (2004).

[6] M D Sumption, M Bhatia, S X Dou, M Rindfliesch, M Tomsic, L Arda, M Ozdemir, Y Hascicek and E W Collings, Supercond. Sci. Technol. 17, 1180 (2004).

[7] Yanwei Ma, Xianping Zhang, Aixia Xu, Xiaohang Li, Liye Xiao, G. Nishijima, S. Awaji, K. Watanabe, Liye Jiao, Ling Xiao, Xuedong Bai, Kehui Wu and Haihu Wen, Supercond. Sci. Technol. 19 (2006) 133.

[8] A. Serquis, L. Civale, D. L. Hammon, X. Z. Liao, J. Y. Coulter, Y. T. Zhu, M. Jaime, D. E. Peterson, F. M. Mueller, V. F. Nesterenko and Y. Gu, Appl. Phys. Lett. 82, 2847 (2002).

[9] Y. Bugolavsky, L. F. Cohen, G. K. Perkins, M. Polichetti, Y. J. Tate, R. Gwilliam and A. D. Caplin, Nature (London) 411, 561 (2001).

[10] P. de Rango, M. Lees, P. Lejay, A. Sulpice, R. Tournie, M. Ingold, P. Germi, and M. Pernet, Nature (London) 349, 770 (1991).

[11] Yanwei Ma, K. Watanabe, S. Awaji, and M. Motokawa, Appl. Phys. Lett. 77, 3633 (2000).

[12] H. Maeda, P. V. P. S. S. Sastry, U. P. Trociewitz, J. Schwartz, K. Ohya, and M. Sato, IEEE Appl. Supercond. 13, 3339 (2003).

[13] A. Yamamoto, J. Shimoyama, S. Ueda, Y. Katsura, I. Iwayama, S. Horii, and K. Kishio, Appl. Phys. Lett. 86, 212502 (2005).

[14] T. C. Shields, K. Kawano, D. Holdom, and J. S. Abell, Supercond. Sci. Technol. 15, 202 (2002).

[15] S. Reich, G. Leitus, and I. Felner, J. Supercond. 15, 109 (2002).

[16] D. Yin, Y. Inatomib, K. Kuribayashi, J. Cryst. Growth 226, 534 (2001).

[17] Yanwei Ma, K. Watanabe, S. Awaji, and M. Motokawa, Phys. Rev. B. 65, 174528 (2002).

[18] S. Tsurekawa, K Harada, T. Sasaki, T. Matsuzaki, T. Watanabe. Mater Trans JIM 41, 991 (2000).

[19] T. Watanabe, Y. Suzuk, S. Tanii, H. Oikawa, Phil. Mag. Lett. 62, 9 (1990).

# Captions

Figure 1 X-ray diffraction patterns of the MgB$_2$ tapes processed in magnetic fields H$_a$ of (a) 0 and (b) 10 T. The data were obtained after peeling off the Fe-sheath. The peaks of MgO are marked by asterisks. The peaks of Fe were contributed from the Fe sheath.

Figure 2 Field dependence of the transport critical current density at 4.2 K for the tapes processed in magnetic fields H$_a$ of zero and 10 T. The measurements were performed in magnetic fields parallel to the tape surface.

Figure 3 Magnetic Jc dependence at 5 K for the tapes processed in magnetic fields H$_a$ of zero and 14 T, as measured by magnetization. The measurements were performed in magnetic fields parallel to the tape surface.

Figure 4 SEM micrographs of samples which were processed in different magnetic fields. (a) 0T tape and (b) 14T tape; (c) 0T bulk and (d) 8T bulk.

Figure 5 Magnetic Jc dependence at 5, 20, and 30 K for the pellet samples processed in magnetic fields H$_a$ of zero and 8 T. The measurements were performed in magnetic fields parallel to the bulk surface.

Table 1   Description of samples used in this work

| Group | Sample type | Sintering temperature/time | Applied field during sintering | Sample surface and field direction during sintering | Tc |
|---|---|---|---|---|---|
| Group I | Fe clad tape | 600°C/1 h | 10 T | Parallel | 35.2 |
|  | Fe clad tape | 600°C/1 h | 0 T |  | 35.5 |
| Group II | Fe clad tape | 700°C/1 h | 14 T | Perpendicular | — |
|  | Fe clad tape | 700°C/1 h | 0 T |  | — |
| Group III | pellet | 800°C/1 h | 8 T | Perpendicular | 36.9 K |
|  | pellet | 800°C/1 h | 0 T |  | 37.1 K |

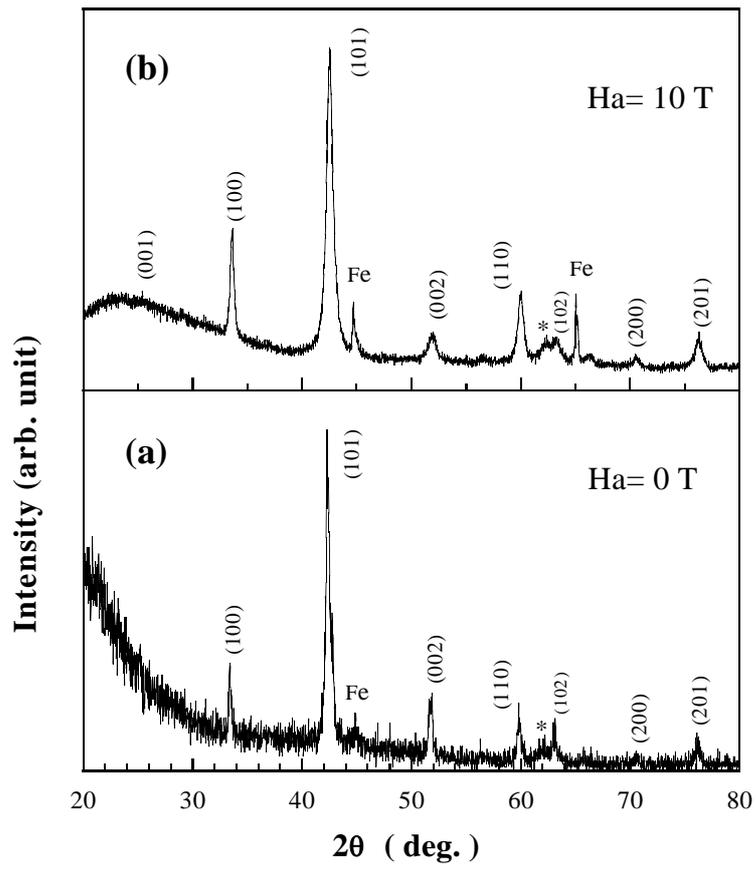

Fig.1 Ma et al.

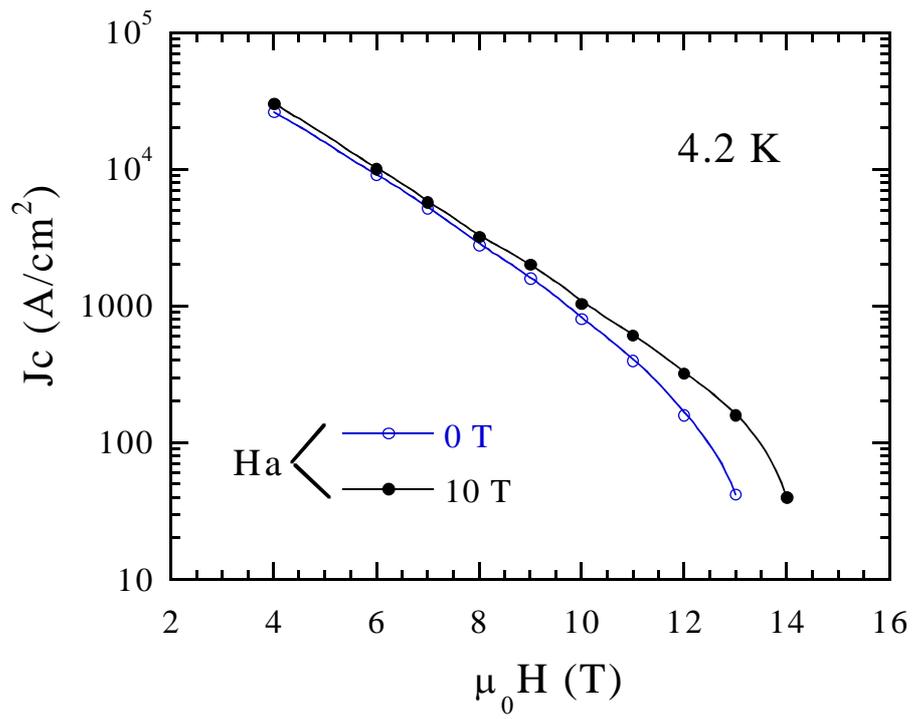

Fig.2 Ma et al.

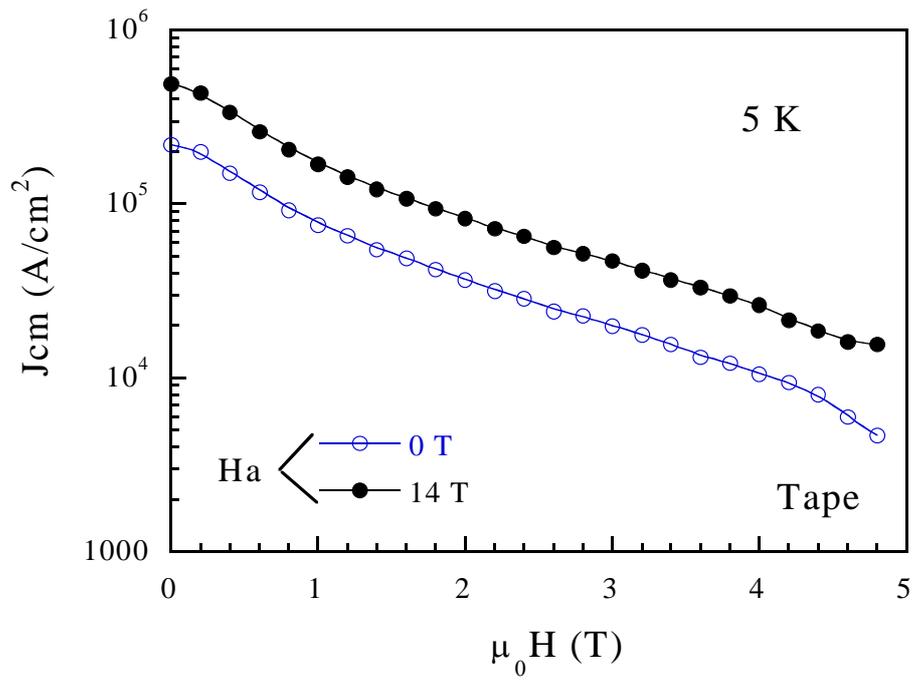

Fig.3 Ma et al.

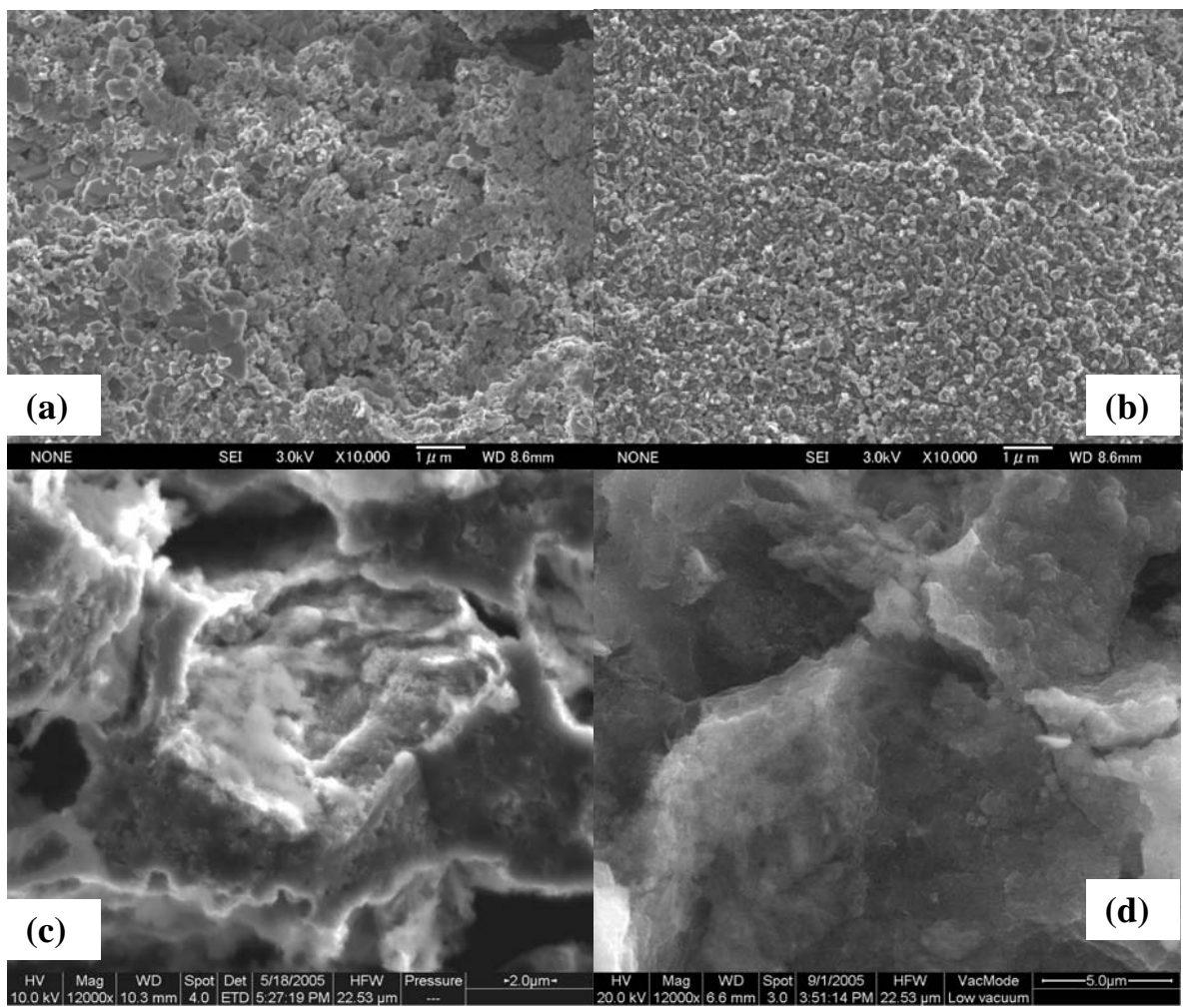

Fig.4 Ma et al.

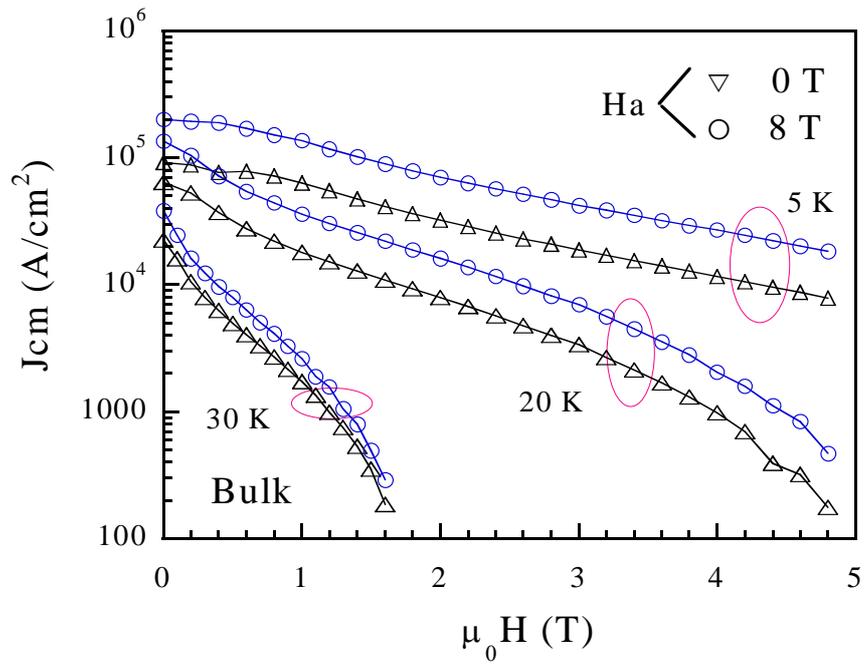

Fig.5 Ma et al.